\documentclass[preprint,3p,11pt]{elsarticle}

\begin{document}

\begin{frontmatter}

\title{Measurement of the atmospheric $\nu_e$ and $\nu_\mu$ energy spectra with the ANTARES neutrino telescope}
\begin{abstract}
This letter presents a combined measurement of the energy spectra of atmospheric $\nu_e$ and $\nu_\mu$ in the energy range between $\sim$100 GeV and $\sim$50 TeV with the ANTARES neutrino telescope.
The analysis uses 3012 days of detector livetime in the period 2007--2017, and selects 1016 neutrinos interacting in (or close to) the instrumented volume of the detector, yielding \textit{shower-like} events (mainly from $\nu_e+\overline \nu_e$ charged current plus all neutrino neutral current interactions) and \textit{starting track} events (mainly from $\nu_\mu + \overline \nu_\mu$ charged current interactions).
The contamination by atmospheric muons in the final sample is suppressed at the level of a few per mill by different steps in the selection analysis, including a Boosted Decision Tree classifier.
The distribution of reconstructed events is unfolded in terms of electron and muon neutrino fluxes. The derived energy spectra are compared with previous measurements that, above 100 GeV, are limited to experiments in polar ice and, for $\nu_\mu$, to Super-Kamiokande.
\end{abstract}

\author[IPHC,UHA]{A.~Albert}
\author[IFIC]{S.~Alves}
\author[UPC]{M.~Andr\'e}
\author[Genova]{M.~Anghinolfi}
\author[Erlangen]{G.~Anton}
\author[UPV]{M.~Ardid}
\author[CPPM]{J.-J.~Aubert}
\author[APC]{J.~Aublin}
\author[APC]{B.~Baret}
\author[LAM]{S.~Basa}
% ...
\author[]{
(The ANTARES Collaboration)
}

\address[IPHC]{\scriptsize{Universit\'e de Strasbourg, CNRS,  IPHC UMR 7178, F-67000 Strasbourg, France}}
\address[UHA]{\scriptsize Universit\'e de Haute Alsace, F-68200 Mulhouse, France}
\address[UPC]{\scriptsize{Technical University of Catalonia, Laboratory of Applied Bioacoustics, Rambla Exposici\'o, 08800 Vilanova i la Geltr\'u, Barcelona, Spain}}
\address[Genova]{\scriptsize{INFN - Sezione di Genova, Via Dodecaneso 33, 16146 Genova, Italy}}
\address[Erlangen]{\scriptsize{Friedrich-Alexander-Universit\"at Erlangen-N\"urnberg, Erlangen Centre for Astroparticle Physics, Erwin-Rommel-Str. 1, 91058 Erlangen, Germany}}
\address[UPV]{\scriptsize{Institut d'Investigaci\'o per a la Gesti\'o Integrada de les Zones Costaneres (IGIC) - Universitat Polit\`ecnica de Val\`encia. C/  Paranimf 1, 46730 Gandia, Spain}}
\address[CPPM]{\scriptsize{Aix Marseille Univ, CNRS/IN2P3, CPPM, Marseille, France}}
\address[APC]{\scriptsize{Universit\'e de Paris, CNRS, Astroparticule et Cosmologie, F-75006 Paris, France}}
\address[LAM]{\scriptsize{Aix Marseille Univ, CNRS, CNES, LAM, Marseille, France }}
% ...
\address[IFIC]{\scriptsize{IFIC - Instituto de F\'isica Corpuscular (CSIC - Universitat de Val\`encia) c/ Catedr\'atico Jos\'e Beltr\'an, 2 E-46980 Paterna, Valencia, Spain}}
% ...
\address[Genova-UNI]{\scriptsize{Dipartimento di Fisica dell'Universit\`a, Via Dodecaneso 33, 16146 Genova, Italy}}
% ...


\begin{keyword}
Neutrino telescope\sep 
Atmospheric neutrinos \sep
ANTARES

\end{keyword}
\end{frontmatter}
\section{Introduction\label{sec:intro}} 
Atmospheric neutrinos are secondary particles produced by cosmic ray (CR) primaries interacting in the Earth's atmosphere.
Due to the need of very large detectors, only a few measurements of the differential flux exist, namely from the AMANDA \cite{Amanda_numu,Amanda_2}, IceCube \cite{IC_40,IC_nue1,IC_nue2,IC_numu1,IC_numu2} and Super-Kamiokande \cite{sk} Collaborations, and a historical measurement from the Frejus Collaboration \cite{Frejus}. The ANTARES Collaboration has reported a measurement of the atmospheric $\nu_\mu$ energy spectrum in \cite{anta_numu}.

Different theoretical frameworks are available to estimate atmospheric neutrino fluxes \cite{honda,honda2,bartol,fluka}. 
At energies from 100 GeV to 1 PeV, the main source of $\nu_\mu$ are semi-leptonic and three-body decays of charged kaons, while the contributions from pion and muon decays dominate below 100 GeV. 
This \textit{conventional} neutrino flux tends towards a power law $\Phi^c_\nu\propto E_\nu^{-\gamma_{CR}-1}$, where $\gamma_{CR}$ is the spectral index of the primary CRs.

Above 100 GeV and up to some tens of TeV, atmospheric $\nu_e$'s come mostly from decays of neutral and charged kaons, and have the same spectral index of conventional $\nu_\mu$. 
Below 100 GeV, $\nu_e$ are predominantly produced by muon decays. 
The $\nu_\mu/\nu_e$ flux ratio is $\sim$2 in the GeV range and increases with energy, reaching a factor of $\sim$20 at 1 TeV. 

At high energies, equal fluxes of $\nu_\mu$ and $\nu_e$ are produced by the decays of charged and neutral D-mesons.
Because of the very short lifetime of these mesons, the resulting flux is called \textit{prompt} neutrino flux \cite{prompt,prompt2} and its energy spectrum, $\Phi^p_\nu\propto E_\nu^{-\gamma_{CR}}$, follows the primary spectrum up to very high energies. The transition from the region in the spectrum dominated by conventional neutrinos to prompt neutrinos is expected to occur at $E_\nu \sim$ 1 PeV for $\nu_\mu$ and around $E_\nu \sim$30 TeV for $\nu_e$.
As a rule of thumb, the primary CR energy is about 20 times higher than the energy of the secondary neutrino. 
Uncertainties on the conventional flux models at neutrino energies above 1 TeV are mainly due to a poor knowledge of primary CR energy spectrum and composition, and of hadronic interactions, in particular of strange quark production mechanisms \cite{honda}. For a recent, detailed description of the hadronic interactions leading to inclusive lepton fluxes, refer to \cite{fedy}.

Finally, $\nu_\tau$ production in the atmosphere is rare: this is dominated by the decay $D^+_s \rightarrow \tau^+ \nu_\tau$, followed by $\tau$ decay. As oscillation effects for atmospheric $\nu$'s are negligible above $\sim$100 GeV, the contribution from tau neutrinos is not considered in this analysis.

This letter describes a strategy to select \textit{shower-like} and \textit{starting track} events (\S \ref{sec:antares}) over the background of atmospheric muons (\S \ref{sec:data}). The distributions of observed events are unfolded (\S \ref{sec:method}) to obtain the energy spectra of both atmospheric $\nu_\mu$ and $\nu_e$, taking into account the detector acceptance (\S \ref{sec:result}). The results are compared with those obtained by other experiments (\S \ref{sec:conc}).
The ANTARES telescope is not able to distinguish between neutrino and antineutrino events. Hence, the unfolded spectra are the sum of $\nu_e+\overline \nu_e$ and of $\nu_\mu+\overline \nu_\mu$, averaged over the zenith region 90$^\circ$--180$^\circ$.

\section{The ANTARES detector and neutrino reconstruction algorithms \label{sec:antares}}

The ANTARES telescope \cite{anta_det} is a deep-sea Cherenkov neutrino detector, located 40 km offshore Toulon, France, in the Mediterranean Sea. The detector comprises a three dimensional array of 885 optical modules \cite{anta_om}, each one housing a 10-inch photomultiplier tube (PMT). 
The optical modules are distributed along 12 vertical strings anchored to the sea floor at distances of about 70 m from each other and at a depth of about 2500 m. The detection of light from upward going charged particles is optimised with the PMTs facing 45$^\circ$ downward.
Particles above the Cherenkov threshold induce a coherent radiation emitted in a cone with a characteristic angle $\theta_C\simeq 42^\circ$ in water.  
For high-energy muons ($E_\mu> 1$ TeV), the contribution of the energy losses due to radiative processes increases linearly with $E_\mu$, and the resulting electromagnetic showers produce additional light. 
Completed in 2008, the telescope aims primarily at the detection of neutrino-induced through-going muons. 

The signals induced in the PMT by detected photons are referred to as \textit{hits} \cite{ars_paper}. The position, time, and collected charge of the hits are used to reconstruct the direction and energy of events induced by neutrino interactions and atmospheric muons. Trigger conditions based on combinations of local coincidences are applied to identify signals due to physics events over the environmental light background due to $^{40}$K decays and bioluminescence \cite{antaresDAQ}.
For astronomy studies, atmospheric muons and atmospheric neutrinos constitute the main source of background. 

This analysis focuses on events induced by neutrinos whose interaction vertices are contained inside (or close to) the instrumented detector volume. These events include: 

\noindent $\bullet$ $\nu_e$ charged current (CC) interactions producing electromagnetic and hadronic cascades, and neutral current (NC) interactions of neutrinos of all flavours inducing hadronic cascades. Due to the radiation and nuclear interaction lengths in water, the cascades extend up to a maximum distance of $\sim$10 m from the interaction vertex, much shorter than the distance between detector strings. These events are thus almost point-like at the scale of the detector and are referred to as \textit{shower-like} events in the following.

\noindent $\bullet$ $\nu_\mu$ CC interactions, with a hadronic cascade near the vertex and a starting muon. Most of these muons are minimum ionising particles, and can travel in water about 4 m per GeV of energy, inducing Cherenkov light over large distances with respect to the interaction vertex position.
These events with a cascade and a track are referred to as \textit{starting track} events in the following.

{
All neutrino candidates used in this letter are selected with an algorithm (denoted in the following as TANTRA) devoted to reconstruct events with interaction vertex inside, or close to, the instrumented volume. 
The initial reconstructed data sample is dominated by downward going atmospheric muons, exceeding by a factor $10^4$ the expected signal of shower-like and starting track events. 
As detailed in \cite{anta_tantra}, the TANTRA likelihood-based method allows reconstructing the vertex coordinates, the neutrino direction, and the neutrino energy, yielding a parameter (a modified $\chi^2$-like quantity) associated to the quality of the fit that is denoted as $M_{est}$.
When the background of atmospheric muons is suppressed, and in dependence of the reconstruction quality $M_{est}$, the neutrino vertex position is determined with a precision up to $\sim$1 m; the neutrino direction is estimated with a median angular resolution of $\sim$$3^\circ$ for a $E_\nu^{-2}$ energy spectrum in the range 1--1000 TeV. Under these assumptions, the uncertainty on the reconstructed neutrino energy can be as low as $\sim$10\% for $\nu_e$.

If arriving in the detector an atmospheric muon, or a muon produced in a neutrino CC interaction with vertex far from the instrumented volume, induces a long sequence of hits characteristic of a long track.}
The track reconstruction algorithm used in off-line ANTARES analyses is called AAFit \cite{anta_aafit} and it is based on a likelihood fit that exploits a detailed parametrisation of the probability density function for the time of the hits. The algorithm provides the track direction with its estimated angular uncertainty and { a proxy for the muon energy loss, which can be used to estimate the parent neutrino energy.
The reconstruction quality is determined by a parameter, referred to as $\Lambda$, which is based on the maximum value of the likelihood fit and the number of degrees of freedom of the fit. The AAFit method is described in \cite{anta_aafit} and in this analysis it is mainly used to remove the largest fraction of atmospheric muons in the data sample. These events are downward going and can be significantly suppressed by a combination of cuts based on the reconstructed track direction and the $\Lambda$ quality parameter, as described in \cite{anta_ps}.}

{ Finally, to improve the rejection of downward going atmospheric muon events, this analysis uses an auxiliary algorithm denoted as GridFit \cite{visser}. GridFit} searches for tracks in 500 different directions covering the full solid angle. The number of hits compatible with a muon track coming from each direction is evaluated and a likelihood fit is performed. 
{ The outcomes of the AAFit and GridFit methods are used in the event selection in order to define cuts allowing to remove (within the available statistic of simulated events) the contamination of atmospheric muons, as detailed in the following section.}


All ANTARES analyses follow a blinding policy to avoid possible biases. The cuts and the selection criteria are studied and optimised on a sample of Monte Carlo (MC) simulated events and only at the end of the full selection chain, these cuts are applied to data. 
A small sample containing 10\% of the real data uniformly distributed over livetime is used to verify the agreement with MC events along the selection.

The simulation chain \cite{anta_MC} starts with the generation of the event and comprises the generation of Cherenkov light, the inclusion of the environmental optical background extracted from real data, and the digitisation of the PMT signals following a \textit{run-by-run} strategy.
This strategy accounts for seasonal variations related to biological activities and for inefficiencies due to the ageing of the PMTs and to biofouling \cite{anta_bio}.

At the end of the full simulation chain, a set of MC files is available for each run of real data, stored in the same format. Simulated files are processed with the same reconstruction algorithms and analysis procedures used for the corresponding data. 
Monte Carlo neutrino events have been generated in the energy range $10 \leq E_\nu \leq 10^8$  GeV, separately for $\nu_e$, $\nu_\mu$ and their antineutrinos, and for CC and NC processes. Details on the simulation chain, hadronic model for cross sections, interaction kinematics and parton distribution functions are given in \cite{anta_MC}.
The same MC sample can be differently weighted to reproduce the conventional atmospheric neutrinos, the prompt neutrinos and theoretical astrophysical signals. 
In the present letter, the atmospheric $\nu_e$ and $\nu_\mu$ fluxes are represented with the same models used in \cite{anta_diffu2}, namely, the conventional component follows the spectrum described in \cite{honda2}, extrapolated at higher energy as in \cite{IC_nue2}, and the prompt contribution as calculated in \cite{prompt}. 
The MC statistics for the atmospheric neutrino sample corresponds to more than two orders of magnitude than for real data.

Finally, for each data run, a file with simulated atmospheric muons (CR$\mu$) is produced with the MUPAGE package \cite{mupage1,mupage2}; in this case, the equivalent MC livetime corresponds to 1/3 of the real run livetime.

\section{Event selection: signal and background\label{sec:data}}

Data collected from 2007 until the end of 2017 have been used. Only runs without high bioluminescence level have been selected. The total livetime corresponds to 3012 days. 
The background is almost entirely due to CR$\mu$'s: after trigger and reconstruction, the expected signal-to-background rate is $\sim$$10^{-4}$. The background suppression is organised in three different steps.

An initial preselection of shower-like and starting track events combines information from both TANTRA and AAFit reconstruction algorithms, according to the following { four} requirements:

\noindent $i)$ the direction of the event as reconstructed by AAFit must be upward going (i.e., zenith angle $>90^\circ$), to reject the largest fraction of CR$\mu$'s;

\noindent $ii)$ the TANTRA's reconstructed event interaction vertex must be contained in a cylindrical volume of axial radius of 300 m and height of 500 m, centred at the centre-of-gravity of the detector modules; 

\noindent $iii)$ the TANTRA estimated angular uncertainty on the event direction must be $< 30^\circ$ and the quality parameter $M_{est}<1000$, to remove poorly reconstructed events;

{ \noindent $iv)$} upward going tracks with AAFit quality parameter $\Lambda>-5.2$ are discarded. This removes muons generated from neutrino interactions outside the detector volume that could have survived previous cuts. These through-going events have already been used in the previous measurement of the $\nu_\mu$ spectrum \cite{anta_numu}. 

After this preselection, the MC signal is reduced by a factor of two with respect to the trigger and reconstruction level, with $\sim$350 survived CR$\mu$'s for each atmospheric neutrino candidate. 

The second step, following \cite{anta_numu,anta_ps,anta_diffu}, uses the AAFit quality parameter, $\Lambda$. The best compromise to suppress the largest percentage of background while keeping a large enough fraction of signal events is obtained by removing events with $\Lambda \le -5.7$. 
After this cut, 25\% of the signal survives, with about 30 remaining background events for each atmospheric neutrino. 
Table \ref{tab:prelim} summarises the number of events passing the preselection and the $\Lambda$ cut for each MC sample. The last row shows the events in the data sample, after unblinding.

\begin{table}[t]
\centering
{\small \begin{tabular}{|l|c|c|}
\hline
       & Preselection & +BDT $>0.33$ \\ 
       & + $\Lambda>-5.7$ &  \\ 
\hline
{ MC} CR$\mu$ & 136700 & $\sim$3 \\
{ MC} Atmospheric $\nu_e$ CC & 242 & 96 \\
{ MC} Atmospheric $\nu_e$ NC & 22 & 9 \\
{ MC} Atmospheric $\nu_\mu$ CC & 3780 & 620 \\ 
{ MC} Atmospheric $\nu_\mu$ NC & 400  & 180 \\
{ MC} Cosmic $\nu$ & 30.4 & 9.2 \\\hline
{ MC total} & 141200 & 917 \\ \hline\hline
Data { (3012 days)} & 133676 & 1016 \\ \hline
\end{tabular} }
\caption{{\small
Number of events in different Monte Carlo samples surviving the preselection and the cut on track quality parameter $\Lambda$ (second column) and the final BDT cut (third column). The last row shows the number of data events in 3012 days of livetime. 
{ 
Cosmic neutrinos are the searched signal in neutrino telescopes; for this analysis they represent an additional background (see \S \ref{sec:method}). The MC Cosmic $\nu$ expectation is computed assuming the flux estimated in \cite{anta_diffu2}.} }}
\label{tab:prelim} 
\end{table}

The final classification of events as signal or background is performed with a Boosted Decision Tree (BDT), defined on a multidimensional parameter space.
A BDT is an algorithm that belongs to the family of supervised machine learning techniques.
To build the classification function, training samples are necessary. 
CR$\mu$ events generated with MUPAGE constitute the background sample; CC+NC interactions of atmospheric $\nu_e$ are used as signal.
{ The CC+NC interactions of $\nu_\mu$ are not used for training the algorithm to reject background.}
This choice is motivated by the fact that the $\nu_e$ flavour produces the cleanest case of shower-like events and it is the most difficult channel to measure in neutrino telescopes. 

For each CR$\mu$ or $\nu_e$ event, the classifier is trained using the following 15 quantities. 
From the TANTRA algorithm, the reconstructed
1) zenith angle and 2) azimuth angle in the local reference frame; 
3) interaction vertex coordinates;  
4) quality parameter estimator, $M_{est}$; 
5) number of detector lines with at least one hit;
6) total number of hits used to reconstruct the event; 
7) angular resolution associated to a shower-like event. 
From the AAFit track-like algorithm, the reconstructed
8) zenith angle and 9) azimuth angle in the local reference frame; 
10) track length inside the detector volume;
11) quality parameter estimator, $\Lambda$;
12) angular resolution associated to a track-like event.
From the GridFit track-like algorithm,
13) the quality parameter;
14) the CR$\mu$ veto parameter, a likelihood variable based on time sequence and charge of the hits in different storeys of the detector, causally-connected under the assumption of a downward going, minimum ionising particle;
15) the number of on-time hits, which assumes that the photons are produced at the Cherenkov angle and arrive at the PMT unscattered.

A ranking of the BDT input variables is derived by counting how often each variable is used to split decision tree nodes, and by weighting each split occurrence by its squared separation gain and by the number of events in the node \cite{classi}. 
None of the variables is found to be significantly dominant; the variable with the highest ranking is the TANTRA zenith angle (1) with score 0.12, followed with the GridFit quality (13) with score 0.10; in the last two positions, the estimators of the angular resolution from TANTRA (8), with score 0.04, and that from AAFit (12), with score 0.023.

\begin{figure}[tbh]
\begin{center}
\includegraphics[width=9.8cm]{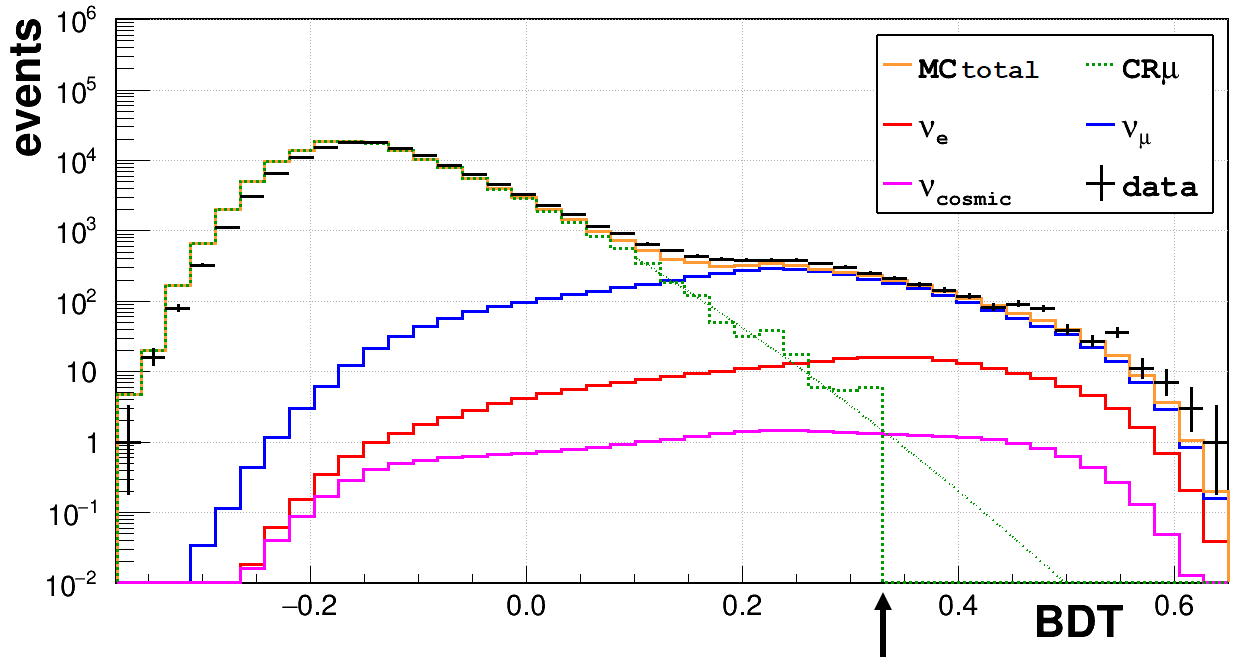}
\end{center}
\caption{{BDT output for events passing the preselection + $\Lambda$ cut. The histograms correspond to different MC samples: training CR$\mu$ (green), training atmospheric $\nu_e$ (red), atmospheric $\nu_\mu$ (blue). The $\nu_\mu$ events are not used for BDT training. The green line corresponds to a Gaussian extrapolation of the CR$\mu$ histogram. 
Both $\nu_e$ and $\nu_\mu$ fluxes include conventional \cite{honda2} and prompt neutrinos \cite{prompt}.
The magenta histogram is the expected contribution from diffuse cosmic $\nu$'s, as parameterised in \cite{anta_diffu2}. The orange histogram is the sum of all MC contributions and the black crosses are real data (3012 days livetime), after unblinding. The BDT cut value is denoted with a black arrow.}}
\label{fig:BDT}
\end{figure}

As shown in Fig. \ref{fig:BDT}, the BDT output is an excellent discriminator between events in the atmospheric $\nu_e$ and background CR$\mu$ samples. The BDT distribution obtained from events induced by atmospheric $\nu_\mu$ CC+NC interactions is also included in the plot. 
As expected, this distribution resembles that of the $\nu_e$ signal. 
The BDT condition $>$0.33 that removes all CR$\mu$'s present in run-by-run MC events is chosen as selection criteria. An extrapolation of the BDT distribution tail, assuming a Gaussian shape, yields a conservative extrapolation of (at most) $\sim$3 background events in the final sample, which are considered in the following.
{ The discrepancy between data and CR$\mu$ for BDT$<-0.2$ is due to the uncertainties in the modeling of the bulk of atmospheric muons in a parameter-space (i.e., direction, energy, muon bundle multiplicity) far from that of neutrino-induced events. The excess of events at large BDT values is consistent with an overall normalization of the MC atmospheric neutrino flux.}

The last column of Table \ref{tab:prelim} shows the number of events in the final sample, after BDT cut. 
{ The neutrino signal is reduced to $\sim$20\% of that present in the previous step, while the background of atmospheric muons is removed from the generated MC sample. Only a maximum residual contamination of 3 events is estimated to survive in the data sample, figure obtained with an extrapolation of the BDT distribution.} As expected, the neutrino signal is still dominated by atmospheric $\nu_\mu$ producing starting tracks: only $\sim$10\% of the selected events originate from $\nu_e$. At 1 TeV, the expected flux ratio is $\Phi_{\nu_\mu}/\Phi_{\nu_e}\sim$20.

\section{Unfolding procedure and detector acceptance\label{sec:method}} 

In order to derive from data the $\nu_e$ and $\nu_\mu$ energy spectra, an unfolding method is used. The two true distributions are deconvolved from the experimentally measured one, based on the best knowledge of the detector and on assumptions made on the interaction rates of the different neutrino flavours. 

In counting experiments, events are grouped into certain regions of phase-space, called \textit{bins}. The main observable quantities in neutrino telescopes are the neutrino direction and energy, which are measured only with finite precision due to inevitable detector effects. 
Consequently, an event may be assigned to a wrong bin. 

The outcome of the unfolding procedure, folded with the detector acceptance and livetime, results in a spectrum that allows a direct comparison with other experiments.
Two major classes of unfolding methods exist: algorithms based on matrix inversion or singular value decomposition, such as the TUnfold \cite{tunfold} algorithm used in this analysis; algorithms based on iterative methods or on the use of Bayes' theorem \cite{dago}. A Bayesian approach has been used, e.g., by the Super-Kamiokande experiment \cite{sk} and in our previous measurement of the $\nu_\mu$ energy spectrum using through-going muons \cite{anta_numu}. For an overview of the commonly used unfolding algorithms, see also \cite{un1,un2}.

The TUnfold algorithm \cite{tunfold} is a widely tested and validated algorithm in the context of high-energy physics and it can handle one or more background sources. The algorithm allows to estimate the number of events in $m$ bins of a $true$ distribution $x_j$, given an $observed$ distribution of $y_i$ in $n$ bins: 
\begin{equation}
y_i = \sum_{j=1}^{m} A_{ij}\cdot x_j + b_i , \quad  1\le i\le n \ ,
\end{equation}
where each bin has a background contribution $b_i$. $A_{ij}$ is a matrix of probabilities describing the migrations from bin $j$ to any of the $n$ bins. The method, interfaced to the ROOT analysis package \cite{root}, uses a least square method with Tikhonov regularisation \cite{tik} and a constraint to fix the total number of events.
The least square minimisation requires a number of degrees of freedom such that $n-m>0$, meaning that the data $y_i$ have to be measured in finer bins than extracted by the unfolding procedure.

The energy estimated by the TANTRA reconstruction algorithm, $E_{\rm reco}$, is used to construct the distribution of $y_i$. 
Events in Fig. \ref{fig:BDT} with BDT$>0.33$ are atmospheric $\nu_\mu$ or $\nu_e$, with a contamination of less than a few per mill from CR$\mu$ and a $\sim$1\% fraction of cosmic neutrinos; both samples are considered as background. 
The unfolding method requires a $(n\times m)$ matrix for the $\nu_\mu$ and $\nu_e$ energies, with $n$ bins of $E_{\rm reco}$ and $m$ bins of true energy $E_{\nu}$. Monte Carlo samples allow the construction of: 

\noindent $\bullet$  $A_{ij}^e$, a $(6\times 3)$ { response} matrix obtained with the simulated samples of $\nu_e$ CC+NC interactions; 

\noindent $\bullet$  $A_{ij}^\mu$, a $(15\times 5)$ { response}  matrix obtained with the simulated samples of $\nu_\mu$ CC+NC interactions. 

The chosen number of bins in ($E_{\rm reco}, E_{\nu}$) for the two samples provides the highest stability in terms of unfolding results applied on MC samples with the same number of events as real data.
In the unfolding procedure, the use of $E_{\rm reco}$ is limited to energies between $\sim$100 GeV and $\sim$50~TeV.
The lower bound is determined by the fact that our reconstruction algorithm cannot reliably reconstruct neutrino energies below 100 GeV. 
Above 50 TeV, the event statistics are significantly reduced by the requirement of the containment of the interaction vertex within, or near to, the instrumented volume. In addition, cosmic neutrinos, whose flux suffers large uncertainties, start to be the dominant ``background''. 

\begin{figure}[tbh]
\begin{center}
\includegraphics[width=10.0cm]{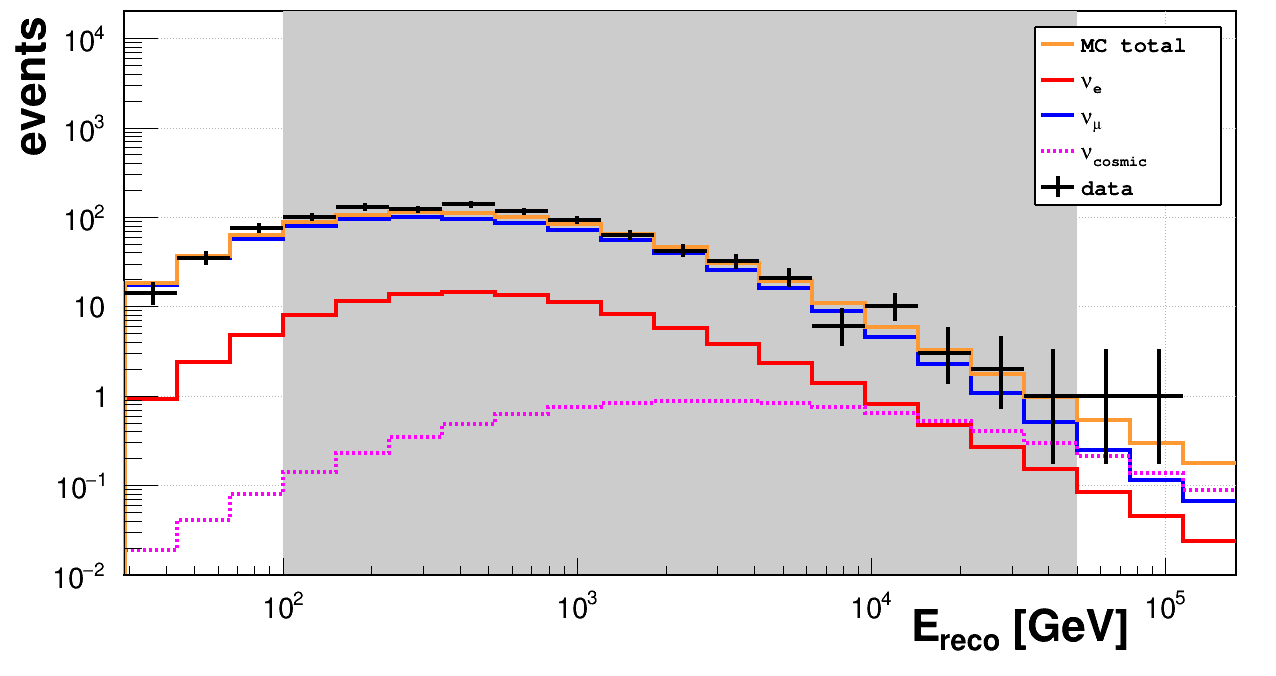}
\end{center}
\caption{{\small Distribution of $E_{\rm reco}$ with the binning used for the construction of the response matrix for the $\nu_\mu$ sample (blue histogram). The red histogram, with the same binning, refers to the $\nu_e$ sample. The magenta histogram is the expected contribution from a cosmic neutrino flux, as estimated in \cite{anta_diffu2}, while the orange histogram includes the sum of all MC contributions. The black crosses correspond to real data. Events in the shaded region are used for unfolding.}}
\label{fig:energy}
\end{figure}

Figure \ref{fig:energy} shows the distribution of $E_{\rm reco}$ for the $\nu_\mu$ sample (blue) with the bin size used for the construction of the $A_{ij}^\mu$ matrix. For completeness, the distribution of $E_{\rm reco}$ for the $\nu_e$ sample (red) is superimposed, although the distribution used for the construction of the $A_{ij}^e$ matrix has a different binning. The expected contribution in the sample of cosmic $\nu$'s of all flavours is also shown.

Concerning the background terms, it includes an extrapolated contribution of 3 track-like CR$\mu$ events. Based on the behaviour of atmospheric muons before the BDT cut, the $b_i^{CR\mu}$ terms are assumed to affect only the $\nu_\mu$ sample, and uniformly in the $E_{\rm reco}$ range. The background from the cosmic neutrino flux (terms $b_i^{c}$), assuming equipartition 
$(\nu_e:\nu_\mu:\nu_\tau) = (1:1:1)$, contributes about equivalently to the $\nu_\mu$ and $\nu_e$ samples, following the $E_{\rm reco}$ distribution shown in Fig. \ref{fig:energy}.

{ The unfolding procedure assumes that in the $\nu_\mu$ distribution there are three background components corresponding to CR's 
($b_i^{CR\mu}$), 50\% of cosmic neutrinos ($b_i^{c}/2$), and the $\nu_e$ fraction ($b^e_i$).  In the $\nu_e$ distribution, there are two background components corresponding to 50\% of cosmic neutrinos ($b_i^{c}/2$) and the $\nu_\mu$ fraction ($b^\mu_i$).}
The algorithm assumes that $\sum_i (b_i^\mu + b_i^e + b_i^{CR\mu} +b_i^{c} )$ is equal to the total number of events.
The events $b_i^\mu$ and $b_i^e$ are assumed to be produced with the fluxes given in \cite{honda2}, as in the default Monte Carlo simulation, with free normalisation. 
Possible variations in their spectral indexes are accounted for in the treatment of systematic effects (see \S \ref{sec:result}).

Table \ref{tab:numuflux} presents the information on the unfolded energy in 5 (3) bins for the $\nu_\mu$ ($\nu_e$) sample. The first two columns contain the energy range of the corresponding bin and the weighted central value of the neutrino energy bin, calculated taking into account the steep decrease of the energy spectrum and the detector response. The third column shows the unfolded number of data events as obtained by the algorithm.

\section{The unfolded energy spectrum\label{sec:result}} 

To transform the unfolded number of events, $N^{\rm evt}$, given in Table \ref{tab:numuflux} into a differential energy flux in the proper units (GeV$^{-1}$ cm$^{-2}$ s$^{-1}$ sr$^{-1}$), the following steps are required:
$i)$ divide each bin by the livetime of 3012 days, obtaining the event rate integrated in the $\log_{10}$ of the neutrino energy over the bin;
$ii)$ divide by the width of the bin (0.54 for $\nu_\mu$ and 0.9 for $\nu_e$); then, transform the $\frac{dN^{\rm evt}}{d\log_{10} E_\nu}$ distribution into the $\frac{dN^{\rm evt}}{dE_\nu}$ one;
$iii)$ divide by the integrated value of the observation solid angle, i.e., $2\pi$ sr;
$iv)$ divide by the detector effective area, $A_{eff}(E_\nu)$, averaged over the { distribution of zenith angles, as reconstructed by the TANTRA algorithm.} 

\begin{figure}[tbh]
\begin{center}
\includegraphics[width=9.8cm]{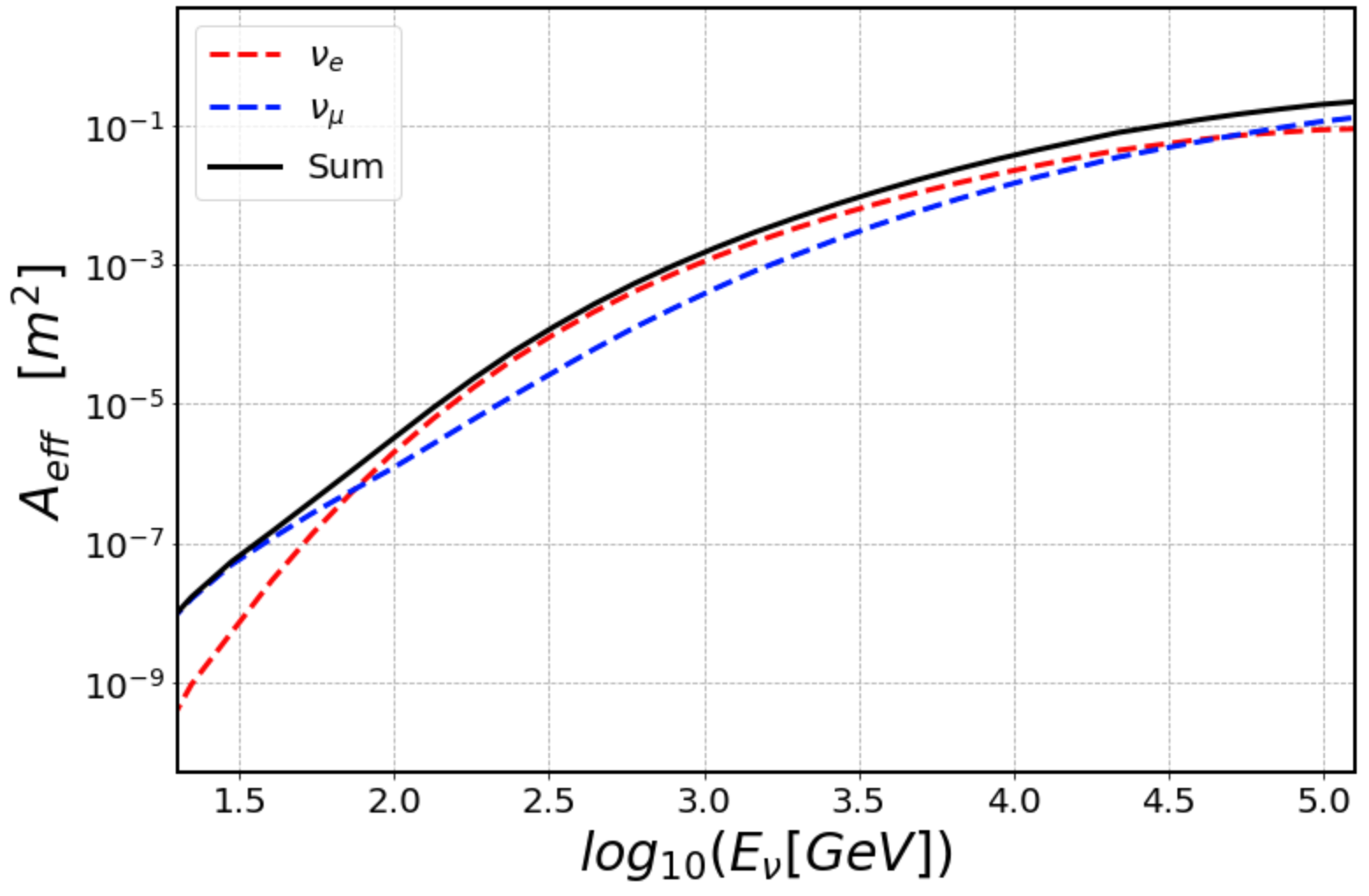}
\end{center}
\caption{{\small Effective area of the ANTARES neutrino telescope for the events with a vertex inside the instrumented volume and selected by the analysis cuts described in this work: $\nu_e$ CC+NC (red line), $\nu_\mu$ CC+NC (blue line). The black solid line is the sum of all the interaction channels and neutrino flavours.}}
\label{fig:aeff}
\end{figure}

The effective area is the figure of merit for a neutrino telescope, representing the size of a 100\% efficient hypothetical target that the detector offers to a certain simulated neutrino flux. It is calculated as 
\begin{equation}
A_{eff}(E_\nu) =\frac{N_{sel}(E_\nu)}{ N_{gen}(E_\nu)}\cdot V_{gen} \cdot  \rho N_A \cdot \sigma(E_\nu) \cdot P_{Earth}(E_\nu), 
\end{equation}
where $N_{sel} (E_\nu)$ and $N_{gen}(E_\nu)$ are, respectively, the number of selected and generated events of a given neutrino energy $E_\nu$ in the generation volume $V_{gen}$; 
$\rho$ and $N_A$ are the matter density and the Avogadro's number; $\sigma(E_\nu)$ is the neutrino cross section; 
$P_{Earth}(E_\nu)$ is the probability of the neutrino to traverse the Earth without being absorbed.
Above 100 GeV, there are no corrections needed for oscillation effects.
Figure \ref{fig:aeff} shows the effective area obtained from the selection of events described in this work.
\begin{table}\centering
\begin{tabular}{c|c|c|c|c|c}
$\Delta \log E_{\nu}$ &  $\overline{\log E_\nu}$  & $N^{\rm evt}$ & $E_\nu^2\Phi_{\nu}$ 
& stat.   & syst. \\ \hline
\multicolumn{6}{c}{Atmospheric muon neutrinos} \\
 \hline
2.00--2.54 & 2.32& 232 & 2.4 $\times 10^{-4}$ & $ \pm $80\% & $\pm$30\% \\
2.54--3.08 & 2.82& 348 & 6.8 $\times 10^{-5}$ & $ \pm $10\% & $\pm$15\% \\
3.08--3.62 & 3.30& 203 &1.4 $\times 10^{-5}$ & $ \pm $15\% & $\pm$15\% \\
3.62--4.16 & 3.80& 58 &2.2 $\times 10^{-6}$ & $ \pm $40\% & $\pm$20\% \\
4.16--4.70 & 4.31& 13 &3.8 $\times 10^{-7}$ & $ \pm $100\% & $\pm$40\% \\ \hline
\multicolumn{6}{c}{Atmospheric electron neutrinos} \\
 \hline
1.9--2.8 & 2.48& 113 & 1.2 $\times 10^{-5}$ & $ \pm $30\% & $\pm$20\% \\ 
2.8--3.7 & 3.08& 21.2& 4.7 $\times 10^{-7}$ & $ \pm $80\% & $\pm$10\% \\
3.7--4.6 & 3.9& 1.4 & 1.7 $\times 10^{-8}$ & $^{+200\%}_{-100\%}$ & $\pm$20\%
\\
\hline
\end{tabular}
\caption{Column 1: bin width $\Delta \log E_\nu \equiv (\log_{10}\frac{E_\nu^{min}}{\textrm{GeV}}$-$\log_{10}\frac{E_\nu^{max}}{\textrm{GeV}}$, where $E_\nu$ is the unfolded neutrino energy. 
Column 2: the weighted centre of the bin, $\overline{\log E_\nu} \equiv \log_{10}\frac{\langle E_\nu\rangle}{\textrm{GeV}}$.
Column 3: the number of unfolded events assigned to the bin, $N^{\rm evt}$.
Column 4: the differential flux (times $E_\nu^2$) computed in the centre of the bin, $E_\nu^2 \Phi_{\nu}$, in units of GeV cm$^{-2}$ s$^{-1}$ sr$^{-1}$.
Columns 5 and 6: the statistical and the total systematic uncertainties, respectively.}
\label{tab:numuflux} 
\end{table}

The fourth column of Table \ref{tab:numuflux} presents the differential flux obtained with the overall procedure. The reported statistical error is determined by the TUnfold method. 

The result of the unfolding process depends on the MC simulation via the construction of the response matrix.
In turn, the simulation depends on a number of parameters with associated uncertainties.
The effects inducing systematic uncertainties on the measurement of the $\nu_\mu$ flux using through-going events have been extensively described in \cite{anta_numu}. { The same systematics affect both the $\nu_\mu$ and $\nu_e$ samples in this analysis and the effects are estimated in dedicated MC simulation datasets (either for $\nu_\mu$ and $\nu_e$), by varying each time only one of the following parameters}:

\noindent $\bullet$ Overall sensitivity of the optical modules, changed by $+10\%$ and $-10\%$. This includes the uncertainty on the conversion of a photon into a photoelectron as well as the angular dependence of the light collection efficiency of each optical module.

\noindent $\bullet$ The uncertainties on water properties, by scaling up and down by $10 \%$ the absorption length of light in water with respect to the nominal value.

\noindent $\bullet$ The uncertainties related to the neutrino fluxes used in the default response matrix of the unfolding procedures, including a slope change of $\pm 0.1$ in the spectral index, independently for $\nu_e$ and $\nu_\mu$.

Each modified MC sample was then used as pseudo-data to construct a new response matrix, used for unfolding. 
The deviation in the content of each $E_\nu$ bin from the spectrum obtained with the default response matrix, $A^e_{ij}$ or $A^\mu_{ij}$, corresponds to the systematic uncertainty associated with the parameter variation. For each energy bin, the total uncertainty is computed as the quadratic sum of each contribution, and the resulting value is reported in the last column of Table \ref{tab:numuflux}.

\section{Results and conclusions\label{sec:conc}}

\begin{figure}[tbh]
\begin{center}
\includegraphics[width=11.0cm]{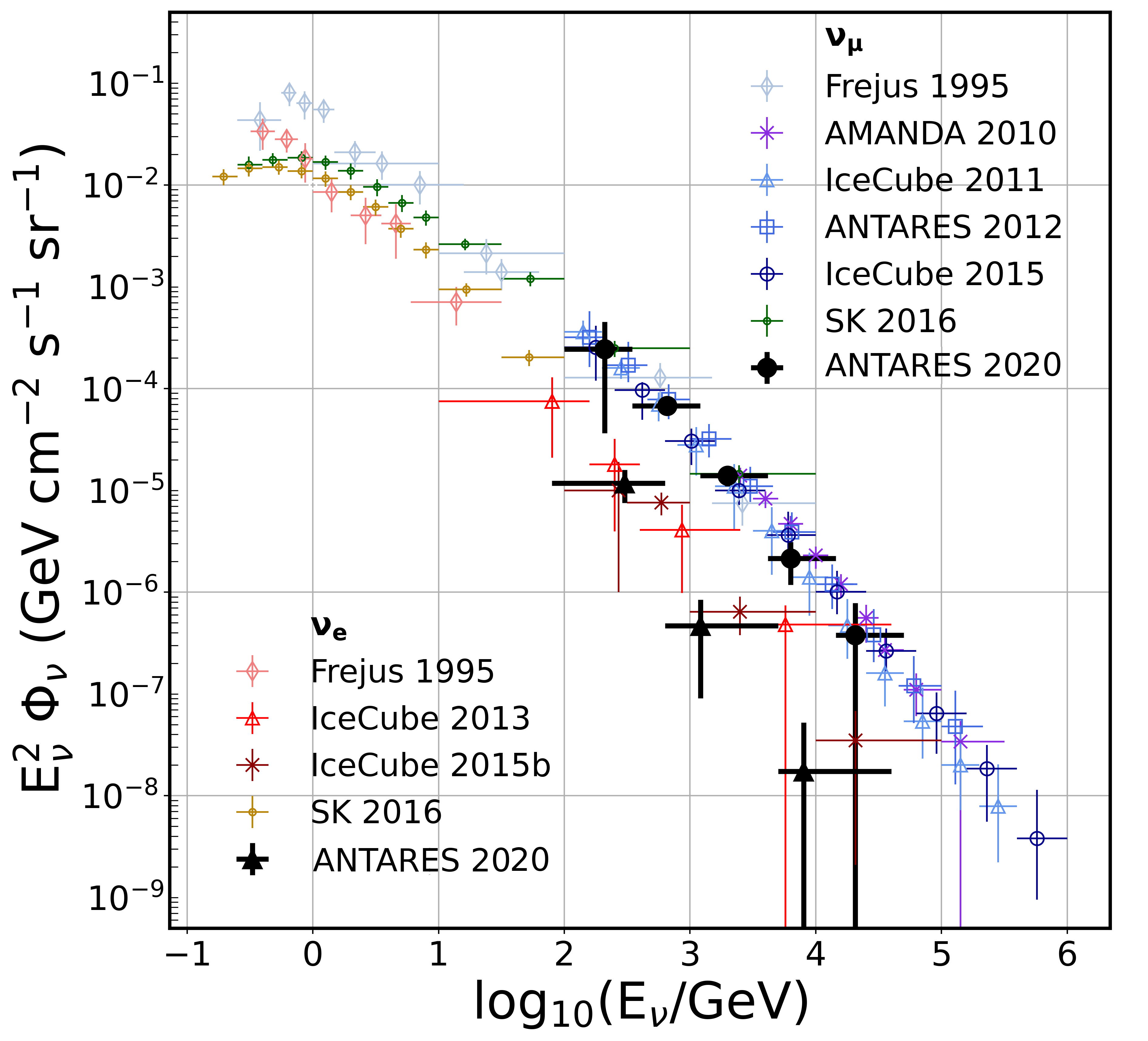}
\end{center}
\caption{{\small Measured energy spectra of the atmospheric $\nu_e$ and $\nu_\mu$ using {\it shower-like} and {\it starting track} events in the ANTARES neutrino telescope (black). The measurements by other experiments (Frejus \cite{Frejus}, AMANDA-II \cite{Amanda_2}, IceCube \cite{IC_numu1,IC_numu2,IC_nue1,IC_nue2}, and Super-Kamiokande \cite{sk}), as well as the previous $\nu_\mu$ flux measurement using a different ANTARES data sample \cite{anta_numu}, are also reported. 
The vertical error bars include all statistical and systematic uncertainties.}}
\label{fig:flux}
\end{figure}

Figure \ref{fig:flux} shows the ($\nu_e+\overline \nu_e$) and ($\nu_\mu+\overline \nu_\mu$) fluxes measured in this work, together with the results from previous experiments. 
Our unfolded atmospheric neutrino spectra, whose statistical errors are largely dominant over the systematic ones, { are 20\%--25\% } below the most recent computations using the SIBYLL-2.3c hadronic interaction model \cite{fedy,fedy2}. 

The measurement of the electron neutrino flux at high energy is challenging, because very large detectors are needed to collect sufficient statistics, and due to large systematic uncertainties.
Each measurement of IceCube-DeepCore \cite{IC_nue1} and IceCube \cite{IC_nue2} rely on about 200 interacting $\nu_e$ in the polar ice medium. The present measurement is performed in seawater, under completely different environmental conditions and systematic uncertainties, yielding consistent results with the ones obtained in polar ice. During a livetime of 3012 days, $\sim$130 $\nu_e$ interactions have been reconstructed within the instrumented ANTARES volume. 
The statistics of the $\nu_e$ sample is not sufficient to test models above a few tens of TeV, where a significant cosmic flux is present and the transition from the conventional to the prompt flux is expected.
Below 100 GeV, the PMT density of the ANTARES detector is insufficient to reconstruct a significant number of events.

Concerning the unfolded $\nu_\mu$ flux, our previous measurement \cite{anta_numu} with a sample of { $\sim$650 through-going events collected in 855 days of livetime and} generated by neutrino interactions external to the instrumented volume almost superimposed the SIBYLL-2.3c model.
The present analysis relies on a totally independent data sample, provided by neutrinos whose reconstructed interaction vertex is inside (or nearby) the instrumented volume of the detector.  
{ The $\nu_\mu$ data sample, summing the $N^{\rm evt}$ events in Tab. \ref{tab:numuflux}, corresponds to $\sim$850 events. 
By scaling the relative livetimes, the number of events in this sample would be $\sim$1/3 of the through-going muons reconstructed as in \cite{anta_numu}. However, due to the superior energy estimate of these (semi)-contained events with respect to the through-going sample, the overall uncertainty on the measured flux is smaller with respect to our previous measurement in the central bins, reaching a precision equivalent to that obtained by the Super-Kamiokande Collaboration in \cite{sk}.}
The present unfolded $\nu_\mu$ flux is more close to that of IceCube with { 40 strings (2011) \cite{IC_numu1} and is 20\%--25\%   below both the flux reported in our previous measurement and the one reported by IceCube using 59 strings (2015) \cite{IC_numu2}}, although consistent within errors.

\vskip 0.5cm
\noindent \textbf{Acknowledgements}
{\small 
The authors acknowledge the financial support of the funding agencies:
Centre National de la Recherche Scientifique (CNRS), Commissariat \`a
l'\'ener\-gie atomique et aux \'energies alternatives (CEA),
Commission Europ\'eenne (FEDER fund and Marie Curie Program),
Institut Universitaire de France (IUF), LabEx UnivEarthS (ANR-10-LABX-0023 and ANR-18-IDEX-0001),
R\'egion \^Ile-de-France (DIM-ACAV), R\'egion
Alsace (contrat CPER), R\'egion Provence-Alpes-C\^ote d'Azur,
D\'e\-par\-tement du Var and Ville de La
Seyne-sur-Mer, France;
Bundesministerium f\"ur Bildung und Forschung
(BMBF), Germany; 
Istituto Nazionale di Fisica Nucleare (INFN), Italy;
Nederlandse organisatie voor Wetenschappelijk Onderzoek (NWO), the Netherlands;
Council of the President of the Russian Federation for young
scientists and leading scientific schools supporting grants, Russia;
Executive Unit for Financing Higher Education, Research, Development and Innovation (UEFISCDI), Romania;
Ministerio de Ciencia e Innovaci\'{o}n (MCI) and Agencia Estatal de Investigaci\'{o}n:
Programa Estatal de Generaci\'{o}n de Conocimiento (refs. PGC2018-096663-B-C41, -A-C42, -B-C43, -B-C44) (MCI/FEDER), Severo Ochoa Centre of Excellence and MultiDark Consolider, Junta de Andaluc\'{i}a (ref. SOMM17/6104/UGR and A-FQM-053-UGR18), 
Generalitat Valenciana: Grisol\'{i}a (ref. GRISOLIA/2018/119) and GenT (ref. CIDEGENT/2018/034) programs, Spain; 
Ministry of Higher Education, Scientific Research and Professional Training, Morocco.
We also acknowledge the technical support of Ifremer, AIM and Foselev Marine
for the sea operation and the CC-IN2P3 for the computing facilities.}









\end{document}